\documentclass[referee]{aa} 
\usepackage{graphicx}

\def\msun{M$_\odot$}
\begin{document}
   \title{Full computation of massive AGB evolution. II.
The role of mass loss and cross-sections}

   \author{P. Ventura
          \and
          F. D'Antona
          }

   \offprints{P. Ventura}

   \institute{Osservatorio Astronomico di Roma
              Via Frascati 33 00040 MontePorzio Catone - Italy\\
              \email{ventura@mporzio.astro.it, dantona@mporzio.astro.it}
             }

   \date{Received ... ; accepted ...}

   \abstract{In the course of a systematic exploration of the uncertainties 
   associated to the input micro- and macro-physics in the modeling of
   the evolution 
   of intermediate mass stars during their Asymptotic Giant Branch (AGB) phase, 
   we focus on the role of the nuclear reactions rates and mass loss. We 
   consider masses 3$\leq$ M/\msun $\leq$ 6.5 for a metallicity 
   typical for Globular Cluster, Z=0.001, and compare the results 
   obtained by computing the full 
   nucleosynthesis with hot bottom burning (HBB), for a network of 30 elements, 
   using either the NACRE or the Cameron \& Fowler (1988) cross-sections.
   The results differ in particular with respect to the $^{23}$Na 
   nucleosynthesis (which is more efficient in the NACRE case) and the 
   magnesium isotopes ratios. For both choices, however, the CNO 
   nucleosynthesis shows that the C+N+O is constant within a factor
   of two, in our models 
   employing a very efficient convection treatment. Different mass loss rates 
   alter the physical conditions for HBB and the length of the AGB phase, 
   changing indirectly the chemical yields. These computations show that the 
   predictive power of our AGB models is undermined by these uncertainties. In 
   particular, it appears at the moment very difficult to strongly accept
   or dismiss that these sources play a key-role in 
   the pollution of Globular Clusters (GCs), and 
   that they have been the main stellar site
   responsible for the chemical anomalies which 
   are observed at the surface of giant and Turn-Off stars of GCs, in the 
   self-enrichment scenarios. 

   }

   \maketitle
%

\section{Introduction} 

Since the pioneering works by Schwarzschild \& Harm (1965, 1967), and Iben 
(1975, 1976), it is now well known that intermediate mass stars (i.e. stars 
with initial masses 1 $\leq$ M/\msun $\leq$ 8, hereinafter IMS) soon after the 
exhaustion of central helium experience a phase of thermal pulses (TPs), during 
which a CNO burning shell supplies for most of the time the global nuclear 
energy release; periodically, a He-burning shell is activated in thermally 
unstable conditions, triggering an expansion of all the outer layers, with the 
consequent extinction of CNO burning (Lattanzio \& Karakas 2001). During the 
AGB evolution these stars suffer a strong mass loss, which ultimately peels-off 
all the envelope mass, leaving a carbon-oxygen compact remnant  
which evolves as a white dwarf. 
The base of the external convective zone of the most massive IMS may become so 
hot ($T_{\rm bce} \geq 30\times 10^6$ K) to favor an intense nucleosynthesis 
(hot bottom burning, HBB), whose results can be directly seen at the surface of 
the star due to the rapidity of convective motions (e.g. Ventura et al. 2002). 

The ejecta of these stars might thus pollute the surrounding medium with 
material which was at least partially nuclearly processed: this is the reason 
why this class of objects has been invoked as a possible explanation of the 
chemical anomalies observed at the surface of giants and turn-off 
globular clusters stars (see e.g. Gratton et al. 2004), in what is commonly 
known as the self-enrichment scenario. An early generation of IMS  
evolved within the first $\sim 100 - 200$ Myr of the cluster life,  
contaminated the interstellar-medium with gas which would be already 
nuclearly processed; this gas might have favored the formation of a later 
generation of stars, which would then show the observed chemical anomalies 
(Cottrell \& Da Costa 1981; D'Antona et al. 1983; Ventura et al. 2001, 2002).

While there is a general agreement that the solution of 
this problem may be looked for in early AGB pollution,
(Gratton et al. 2004), the quantitative agreement between
the models and the abundance patterns shown by GC stars
is not good (Denissenkov \& Herwig 2003; 
Denissenkov \& Weiss 2004). 
On the other hand, the AGB evolution of these stars is
found to be strongly dependent on the convective model
which is used to find out the temperature gradient within
the external convective zone (Renzini \& Voli 1981;
Bl\"ocker \& Schonberner 1991; Sackmann \& Boothroyd 1991;
D'Antona \& Mazzitelli 1996; Ventura \& D'Antona 2005,
hereinafter paper I).
The chemical content of their ejecta, in particular 
for some key-elements which are anticorrelated 
like oxygen and sodium, and magnesium and aluminum, 
is strongly dependent 
not only on convection, but also on the assumed mass 
loss rate and on the nuclear reaction rates.
At the moment these uncertainties seriously undermine
the predictive power of AGB models, and thus limit the
predictions which can be made concerning their role
within the framework of the self-enrichment scenario.

We investigate the AGB evolution of initial
masses 3 $\leq$ M/\msun $\leq$ 6.5, and 
focus our attention on their main physical 
properties, and on the chemical content of 
their ejecta. In paper I we explored the 
dependence of the results
on the convective model. In this work we complete
the exploration by investigating the sensitivity of 
the results on: 1) the
nuclear cross sections: we compare two sets
of models calculated by assuming the Angulo et al.
(1999) NACRE cross sections and those by Caughlan
\& Fowler (1988, hereinafter CF88); 2) the mass loss
rate.

\section{The evolutionary code}
The stellar evolutions discussed 
in this paper were calculated by 
the code ATON2.1, a full description of which can be found in 
Ventura et al. (1998) (ATON2.0 version).
The latest updates of the code, concerning the nuclear network,
are given in paper I. 
The interested reader may find on the 
afore mentioned papers a detailed description 
of the numerical structure of the code, and 
of the macro and micro-physics which 
is used to simulate the stellar evolutions. 

\subsection{Convection}
The code allows us to calculate the temperature gradient
within instability regions either by adopting the
traditional MLT (Vitense 1953; B\"ohm-Vitense 1958), 
or the FST model (Canuto \& Mazzitelli 1991; 
Canuto et al. 1996) for turbulent convection. 
The interested reader may find in 
Canuto \& Mazzitelli (1991) a detailed 
description of the physical differences between 
the two models.

As we shall see, during the AGB evolution a non negligible 
fraction of the global nuclear release is generated within 
the convective envelope, therefore it is mandatory 
to adopt a diffusive approach, treating simultaneously 
mixing and nuclear burning.
We therefore solve for each element the diffusion equation 
(Cloutman \& Eoll 1976):

\begin{equation}
$$
  \left( {dX_i\over dt} \right)=\left( {\partial X_i\over \partial t}\right)_{nucl}+
  {\partial \over \partial m_r}\left[ (4\pi r^2 \rho)^2 D {\partial X_i
  \over \partial m_r}\right]  \label{diffeq}
$$
\end{equation}

\noindent
stating mass conservation of element $i$. The diffusion 
coefficient $D$ is taken as $D={1\over 3}ul$,
where $u$ is the convective velocity and $l$ is 
the convective scale length.
We allow velocity to decay exponentially starting from the
formal convective boundaries as:

\begin{equation}
$$
u=u_b exp \pm \left( {1\over \zeta f_{thick}}ln{P\over P_b}\right) 
$$
\end{equation}

\noindent
where $u_b$ and $P_b$ are, respectively, turbulent 
velocity and pressure at the convective boundary, P 
is the local pressure, $\zeta$ a free parameter 
connected with the e-folding distance of the decay, 
and $f_{thick}$ is the thickness of the convective 
regions in fractions of $H_p$.  

A detailed description of the treatment of convective velocities in the 
proximity of the formal borders of the convective zones (fixed by the 
Schwarzschild criteria) can be found in Sect.2.2 of Ventura et al.(1998). 
In the same paper (Sect.4.2) the interested reader may also find an
extensive discussion on the extra-mixing determined by the use of a 
non-zero $\zeta$.

The models presented in this paper adopt the FST convection, 
and the parameter $\zeta$ is fixed at $\zeta=0.02$. 
No extra-mixing has been assumed from the
base of the convective envelope: therefore 
the extension of the various dredge-up episodes, 
and the consequent changes of the surface 
chemical composition, must be considered 
as lower limits.

\subsection{Nuclear network}
The nuclear network includes 30 chemical 
species up to $^{31}P$ and 64 nuclear reactions. 
The list of all the reactions included in the 
nuclear network can be found in paper I.
The relevant cross-sections are taken either from
Caughlan \& Fowler (1988, CF88) or from Angulo et al 
(1999, NACRE). In the range of
temperatures which are of interest here 
($7.5 \leq \log(T) \leq 8.2$, which are
the typical values at the base of 
the external convective zone of massive AGBs) 
the largest differences between the two sets of 
cross-sections are the following:
\begin{itemize}
\item
{$^{17}$O destruction by proton fusion
is achieved much more easily in the NACRE case;
also, contrary to CF88, the 
reaction $^{17}$O(p,$\alpha$)$^{14}$N is favored
with respect to $^{17}$O(p,$\gamma$)$^{18}$F.}

\item{The cross-sections of the reaction 
$^{22}$Ne(p,$\gamma$)$^{23}$Na are larger by
$\sim$ 3 orders of magnitude in the NACRE
case, which makes sodium production much 
easier. As for sodium burning by proton
fusion, the channel leading to the formation
of $^{24}$Mg is favored in the NACRE case.}  

\end{itemize}

\subsection{Mass loss}
The mass loss rate is calculated according to Bl\"ocker (1995),
who modifies the Reimer's formula in order to simulate the
strong mass loss suffered by these stars as they climb
along the AGB. The complete expression is:
\begin{equation}
$$
\dot M=4.83\times 10^{-9} M^{-2.1} L^{2.7} \dot M_R
$$
\end{equation}
\noindent
where $\dot M_R=10^{-13}\eta_R LR/M$ is the canonical Reimer's
rate, and $\eta_R$ is a free parameter directly connected with 
the mass loss rate. 

$\dot M$ was described according to Eq.3,
with the parameter $\eta_R=0.02$ for the ``standard''
case; we then consider evolutions with $\eta_R=0.1$ 
and $\eta_R=0.2$. In all cases mass loss was applied 
for all the evolutionary phases.

\subsection{Model inputs}
We evolved models with initial masses $3M_{\odot}
\leq M \leq 6.5M_{\odot}$ starting from the pre-main
sequence along the whole TPs phase. Above the upper
mass limit, models ignite carbon
in the center, skipping the AGB phase. Below the
lower limit models do not achieve HBB conditions.

When the envelope mass becomes ``small'' 
($M_e < 1M_{\odot}$) a much higher temporal resolution 
is required, which renders the computations extremely 
time-consuming; since the chemical yields are almost 
unaffected by the following phases, we decided to stop 
the evolutions when the mass of the envelope falls below 
$\sim 0.5M_{\odot}$. 

We adopted an initial metallicity, $Z$, 
typical of those globular clusters (GCs) like NGC6752,
M3, M13, whose stars show the largest chemical 
anomalies, i.e. $Z=0.001$ and $Y=0.24$. For all
the elements included in our network, we adopted
solar-scaled initial abundances. This is to be taken into
account if we want to compare the results with observations,
as the starting initial mass fractions 
of abundant elements play a role
in the determination of the final yield.
E.g., as [O/Fe]$\sim +0.3$ in population II stars, this 
initial abundance will be remembered in the evolution 
with oxygen depletion. Numerical tests we pursued show 
that the results for oxygen may be roughly scaled up by 
the initial enhancement with respect to the solar scaled 
value. For instance, if the solar scaled model produces a 
yield with [O/Fe]=--0.5, the yield starting from initial 
[O/Fe]=+0.3 would have been [O/Fe]=--0.2.

\section{NACRE results}
\subsection{The pre-AGB phase}
Table~\ref{physics} summarizes the main physical 
properties of our models, related to the evolutionary 
phases preliminary to the AGB evolution.
During the main sequence phase the models develop a central
convective region which progressively shrinks in mass,
with a maximum extension ranging from $\sim$ 0.72\msun\
for the 3\msun\ model up to $\sim$2\msun\
for the 6.5\msun\ model.

\begin{figure}
\caption{The duration of the phases of hydrogen and 
    helium burning as a function of the initial
    mass for the NACRE intermediate mass models.
    {\bf Top}: H-burning time; {\bf Middle}:
    He-burning time; {\bf Bottom}: Ratio between
    the He-burning and the H-burning times.}
         \label{times}
\end{figure}

The total duration of the H-burning phase (t(H))is a decreasing 
function of mass: the less massive model, with initial
mass M=3\msun, consumes central hydrogen in $\sim 275$
Myr, while the 6.5\msun\ model keeps burning hydrogen
for 54 Myr (fig.~\ref{times}, top panel).

   \begin{table}
      \caption[]{Physical properties of the NACRE models.}
         \label{physics}
     $$ 
           \begin{array}{c c c c c c c c}
            \hline
            \noalign{\smallskip}
M  &  t(H)^a  & M_{c,H}^b & t(He)^a &  M_{c,He}^c  & M_{1dup}^d  & 
M_{2dup}^e    & \delta(M_{2dup})^f \\
            \noalign{\smallskip}
            \hline
            \noalign{\smallskip}
            3.0 & 276 & 0.72 & 56  & 0.28 & 0.85 & 0.78 & 0.01     \\
            3.5 & 195 & 0.93 & 33  & 0.32 & 1.19 & 0.84 & 0.05     \\
            4.0 & 145 & 1.04 & 23  & 0.36 & 1.50 & 0.87 & 0.15     \\
            4.5 & 112 & 1.21 & 16  & 0.44 & 1.86 & 0.90 & 0.26     \\
            5.0 & 90  & 1.40 & 12  & 0.48 & 2.10 & 0.93 & 0.37     \\
            5.5 & 75  & 1.57 & 10  & 0.57 & 2.42 & 0.97 & 0.48     \\
            6.0 & 63  & 1.85 & 7.6 & 0.62 & 2.70 & 1.00 & 0.58     \\
            6.5 & 54  & 2.05 & 6.3 & 0.70 & 2.85 & 1.05 & 0.68     \\

            \noalign{\smallskip}
            \hline
         \end{array}
     $$ 
\begin{list}{}{}
\item[$^{\mathrm{a}}$] Times are expressed in Myr.
\item[$^{\mathrm{b}}$] Maximum extension (in $M_{\odot}$) 
of the convective core during H-burning.
\item[$^{\mathrm{c}}$] Maximum extension (in $M_{\odot}$) 
of the convective core during He-burning.
\item[$^{\mathrm{d}}$] Mass coordinate (in $M_{\odot}$) 
of the innermost layer reached during the first dredge-up.
\item[$^{\mathrm{e}}$] Mass coordinate (in $M_{\odot}$) of 
the innermost layer reached during the second dredge-up.
\item[$^{\mathrm{f}}$] Amount (in $M_{\odot}$) of  
dredged-up material nuclearly processed by CNO burning.

\end{list}
   \end{table}

During the first dredge-up, following H-exhaustion, the
convective envelope reaches stellar layers which were
previously touched by CNO burning via the CN cycle, with
the conversion of some $^{12}$C to $^{14}$N:
consequently, the surface $^{14}$N is increased by
a factor of $\sim 2$, while $^{12}$C decreases from
the initial value of X($^{12}$C)=1.73 $\times 10^{-4}$
to X($^{12}$C)=1.3 $\times 10^{-4}$. The lithium 
surface abundance drops by a factor of $\sim 50$,
because surface lithium is mixed within an
extended region where lithium was previously destroyed
via proton fusion; this drop is dependent
on the stellar mass, and a spread of a factor 
of $\sim 2$ is found among the models.

During the following phase of core helium burning
a convective core is formed, again with a dimension
increasing with mass: it is $\sim 0.28M_{\odot}$ for
the $3M_{\odot}$ model, while it is $\sim 0.7M_{\odot}$
for the $6.5M_{\odot}$ model (see the 5th column of
Table~\ref{physics}).
We see from Fig.~\ref{times} (bottom panel)
that the ratio between the He-burning (t(He))
and the MS times is decreasing with mass, ranging from
$\sim 20\%$ to slightly higher than $10\%$ for the
6.5M$_{\odot}$ model. Once helium is burnt-out in 
the stellar core, $3\alpha$
reactions carry on in an intermediate layer, triggering
a general expansion of the structure, which eventually
extinguishes the CNO burning shell. The general cooling
of the star favors the formation of a very deep and
extended external convective zone, in what is commonly
known as the second dredge-up episode.

\begin{figure}
\caption{{\bf Top}: The mass coordinate of the
    innermost point reached by the base of the envelope
    during the second dredge-up for the same models
    discussed in fig.~\ref{times}. {\bf Bottom}:
    The penetration (in solar masses) of the base
    of the external convective zone beyond the location
    of the CNO burning shell during the second 
    dredge-up.}
         \label{2dup}
\end{figure}

\begin{figure}
\caption{The variation of the surface chemical abundances
        of some elements following the 2nd dredge-up.
       {\bf Top}: Helium mass fraction; {\bf Middle}:
       Variation of the $^{16}$O abundance, expressed
       as the logarithm of ratio between the final 
       abundance and the initial mass fraction; {\bf
       Bottom}: The same as the middle panel, but for
       $^{23}$Na.}
         \label{2dupchem}
\end{figure}

From the top panel of fig.~\ref{2dup} we see that the mass coordinate
corresponding to the maximum penetration of the outer
envelope is slightly increasing with mass, with a
difference of $\sim 0.2$\msun\ between the 3
and 6.5$M_{\odot}$ models. The most interesting quantity 
is however shown in the bottom panel of the same
figure, where we report the variation as a function of
the initial mass of $\delta M=M_{CNO}-M_{min}$, where 
$M_{CNO}$ and $M_{min}$ are, respectively, the location of
the CNO burning shell immediately before the second 
dredge-up, and the minimum point (in mass) reached by the
base of the outer convective zone. $\delta M$ is therefore
a measure of the amount of processed material which is
carried to the surface during the second dredge-up. 
We see that a poor mixing is expected in the 3$M_{\odot}$
model, while in the 6.5$M_{\odot}$ case $\sim 0.7$M$_{\odot}$
of CNO processed material is mixed with the surface layers.
During the second dredge-up the surface $^{14}$N is 
increased by another factor of $\sim 2$, the carbon abundance
decreases to X($^{12}$C) $\sim 1.15 \times 10^{-4}$, while
lithium is not dramatically affected, because the surface
lithium abundance was already heavily lowered during the
first dredge-up. At the second dredge-up, the
helium, sodium, and oxygen 
abundances are changed (depending on the stellar mass), 
as can be seen in the three panels of fig.~\ref{2dupchem}. 
This can be understood on the basis of the following considerations:
\begin{itemize}
\item{The amount of mixed material during the second
dredge-up sensibly increases with the stellar mass
(see fig.~\ref{2dup}).}
\item{During the second dredge-up the base of the 
outer envelope reaches layers which were previously 
touched by full CNO burning, thus explaining the
oxygen reduction.}
\item{The amount of material previously touched
by CNO burning and mixed to the surface is
larger than in the first dredge-up case, thus
the increase of the surface helium abundance
is very large, especially in the most massive 
models.} 
\end{itemize}

\begin{figure}
\caption{The maximum luminosity ({\bf top}) and temperature
        at the base of the convective envelope ({\bf bottom})
        achieved by the standard NACRE models
        during their AGB evolution.}
         \label{agbmax}
\end{figure}

\subsection{The AGB phase}
For most of the AGB evolution the global nuclear energy 
release is generated within a CNO burning shell, which 
may also overlap, in some cases, with the external 
convective zone. As hydrogen is consumed, the core mass
increases, and the CNO burning takes place at higher
temperatures, thus favoring an increase of the
stellar luminosity. This is halted by mass loss, 
which progressively reduces the mass of the envelope,
and eventually leads to a general cooling of the
outer stellar layers. The base of the convective zone
becomes cooler and cooler in the latest evolutionary 
stages; when the mass of the envelope drops below
$\sim$ 1M$_{\odot}$ the temperatures within the whole
external zone become so small to prevent any further
nucleosynthesis. For each model, we may therefore
find out maximum values of both luminosity and
$T_{\rm bce}$, which we report as a function of the initial
mass of the star in the two panels of fig.~\ref{agbmax}.
We see that even the least massive of our models, i.e.
the 3M$_{\odot}$ model, achieves at the base of the
external convective zone temperatures so large
($T_{\rm bce} \sim 7.75 \times 10^7$ K) to trigger HBB.

In all the models, shortly after the
beginning of the TPs phase, the TDU operates
following each TP, changing the
surface chemistry. The efficiency $\lambda$ of 
the TDU increases with the evolution, and is 
higher in the less massive models. We find that
for $M \geq 5M_{\odot}$ a maximum value of
$\lambda \sim 0.4$ is attained in the latest
evolutionary stages, while a significantly
larger value of $\lambda \sim 0.7-0.8$ is
reached along the evolutions of the models
with masses $3-3.5M_{\odot}$.

The chemical composition of the ejecta of our NACRE 
models, calculated over the lifetime of the star,
is summarized in Table~\ref{ejecta}. With 
the only exception of lithium, for the other 
elements we indicate the logarithm of the ratio 
between the average chemical
abundance of the ejecta and the initial value. 
Therefore, a value of 0 indicates that the chemical
content of the ejecta is the same as the initial
chemistry.

\begin{figure*}
\centering{
           }

\caption{Variation of the surface chemical abundances
   of the CNO elements during the evolution of the
   same models as in fig.~\ref{times}. For clarity reason,
   for $^{12}$C we decided to show only the variation 
   at the surface of the models with initial masses
   3,4,5 and 6M$_{\odot}$.}
   \label{cnonacre}
\end{figure*}

\subsection{The CNO elements}
In the three panels of Fig.~\ref{cnonacre} we show the variation
during the AGB evolution (including also the changes
due to the second dredge-up) of the surface abundances
of the CNO elements; we chose the stellar mass as abscissa
in order to have an idea of the average chemical content 
of the ejecta of these stars. 
We see (left-lower panel) that oxygen is depleted in 
all cases apart from the 3M$_{\odot}$ model, 
the heaviest depletion being for the largest masses,
in agreement with the physical situation present at the
base of the external convective zone. 
We note the apparently anomalous behavior 
of the 6 and 6.5M$_{\odot}$ models, which have a flatter
declining profile of the surface oxygen with mass: this
can be understood on the basis of the fact that these
latter models are already extremely luminous during the
pre-AGB phase, so that a large fraction of the mass is lost
when the oxygen abundance is still close to the value
left behind by the second dredge-up.

In the right panel we may follow the evolution of the
surface abundance of carbon: we note an early phase of
destruction in all models, which corresponds to the stage when
only the CN cycle is active, followed by a later increase, 
when the full CNO cycle is activated; in the less massive 
models we recognize the signature of an efficient third 
dredge-up. 

   \begin{table*}
      \caption[]{Chemical content of the ejecta of the NACRE models.}
         \label{ejecta}
     $$ 
           \begin{array}{c c c c c c c c c c c c c}
            \hline
            \noalign{\smallskip}
             M_{ZAMS} & Y^a & \log(\epsilon(^7Li))^b & [^{12}C]^c & [^{14}N] & [^{16}O] & (C+N+O)^d & [^{23}Na] & [^{24}Mg] & [^{25}Mg] & [^{26}Mg] & ^{25}Mg/^{24}Mg & ^{26}Mg/^{24}Mg   \\
            \noalign{\smallskip}
            \hline
            \noalign{\smallskip}
            3.0 & 0.26 & 2.41 & -0.10 & 1.52 &  0.01 & 3.24 & 1.21 & 0.03 & 0.22 & 0.56 & 0.20  & 0.50   \\
            3.5 & 0.27 & 2.20 & -0.44 & 1.37 & -0.19 & 2.18 & 0.75 &-0.01 & 0.37 & 0.40 & 0.31  & 0.39   \\
            4.0 & 0.29 & 2.06 & -0.56 & 1.28 & -0.39 & 1.70 & 0.46 &-0.28 & 0.60 & 0.41 & 0.98  & 0.73   \\
            4.5 & 0.32 & 1.96 & -0.60 & 1.24 & -0.49 & 1.51 & 0.02 &-0.66 & 0.60 & 0.38 & 2.37  & 1.63   \\
            5.0 & 0.32 & 1.91 & -0.70 & 1.13 & -0.61 & 1.17 &-0.16 &-0.95 & 0.50 & 0.35 & 3.68  & 3.00   \\
            5.5 & 0.32 & 1.94 & -0.80 & 1.04 & -0.59 & 0.99 &-0.37 &-1.13 & 0.53 & 0.31 & 6.13  & 4.18   \\
            6.0 & 0.32 & 1.91 & -0.82 & 1.01 & -0.54 & 0.96 &-0.43 &-1.23 & 0.60 & 0.28 & 8.96  & 4.96   \\
            6.5 & 0.32 & 2.25 & -0.81 & 1.01 & -0.46 & 1.00 &-0.37 &-1.27 & 0.66 & 0.25 & 11.34 & 5.02   \\

            \noalign{\smallskip}
            \hline
         \end{array}
     $$ 
\begin{list}{}{}
\item[$^{\mathrm{a}}$] Helium mass fraction.
\item[$^{\mathrm{b}}$] $\log(\epsilon(^7Li))=\log(^7Li/H)+12.00$
\item[$^{\mathrm{c}}$] $[A]=\log(X(A)_{ej})-\log(X(A)_{in})$.
\item[$^{\mathrm{d}}$] Ratio between the average (C+N+O) abundance of
the ejecta and the initial (C+N+O) value.
\end{list}
   \end{table*}

Even in this case we note the peculiar behavior
of the 3$M_{\odot}$ model, in which the 3rd dredge-up
is highly efficient since the first TPs, so that the
surface $^{12}$C abundance increases up to
$\log$(X($^{12}$C))$\sim -3.4$; only in a later time,
$\sim 200,000$ yr after the beginning of the AGB 
phase, when $\sim 0.5$M$_{\odot}$ have been lost,
HBB occurs, and the $^{12}$C abundance starts to
decrease. The 3rd dredge-up after 10 TPs is so
efficient that also some $^{16}$O is carried outwards;
$^{16}$O reaches a maximum abundance after 15 TPs 
and then decreases approximately to the initial 
value: this is the only model for which we find an 
oxygen content of the ejected material which is 
larger than the original composition (see the 6th 
column of tab.~\ref{ejecta} and the top panel of 
Fig.~\ref{yield1} ).
The combination of the effects of HBB and of
the almost constant surface abundance of $^{16}$O
prevents the formation of a carbon star.
From this discussion we argue that 3M$_{\odot}$ is 
approximately the lower limit for models achieving HBB 
with the full CNO cycle operating during the AGB 
evolution. No $^{16}$O depletion can be achieved 
in less massive models.

The carbon, nitrogen and oxygen abundance of the ejecta 
are reported in columns 4-6 of Table~\ref{ejecta}.
We see that the carbon content of the expelled material
is always smaller than the initial value, due to the drop 
of the surface carbon which follows the first and 
especially the second dredge-up. The depletion factor
is lower the lower is the mass, because in the less
massive models more carbon is produced later in the
AGB evolution by the 3rd dredge-up. We note that even for
the 3M$_{\odot}$ model, despite the early phase of
$^{12}$C production at the beginning of the TPs phase
(see the right panel of fig.~\ref{cnonacre}), we find 
a negative $[^{12}$C], due to a later phase of 
$^{12}$C depletion at the base of the convective envelope
via proton fusion.

From column 5 of tab.~\ref{ejecta} we see that the
nitrogen abundance of the ejecta is always at least
a factor of $\sim 10$ larger than the initial value.
Nitrogen is mixed to the surface through 
the first and second dredge-up, then its 
surface abundance increases due to HBB, via CN and
ON cycling, and may further increase following each third 
dredge-up episode, via the conversion of additional primary 
$^{12}$C mixed into the envelope.
Also in this case, it is the higher number of TPs 
and the larger efficiency of 3rd dredge-up 
episodes the reason of the larger $^{14}$N 
abundances found in the ejecta of the less 
massive models.

As for oxygen, with the only exception of the
3M$_{\odot}$ model we find, in agreement with what 
is shown in the left-lower panel of fig.~\ref{cnonacre},
$[^{16}$O]$<0$ in all cases (see also the top
panel of fig.~\ref{yield1}). We note that the minimum
value of $[^{16}$O], i.e. $[^{16}$O]=--0.61, is reached 
for M=5$M_{\odot}$, because the yield of more massive 
models is influenced by the very strong mass loss 
already efficient at the 2nd dredge-up, when no HBB 
had started yet.

The oxygen isotopes show a similar behavior in all
our models. The surface abundance of $^{17}$O is increased
during the second dredge-up by $\sim 0.2$ dex; at the
very beginning of the AGB evolution $^{17}$O is produced
at the base of the external zone due to partial
$^{16}$O burning, so that, particularly in the most
massive models, its abundance is increased by a factor
of $\sim 10$. Later on, when $^{16}$O burning is more 
efficient, the surface $^{17}$O abundance reaches an
equilibrium value, and then decreases as the $^{16}$O.
The maximum surface $^{17}$O abundance is log[X($^{17}$O)]
$\sim -5.5$ for the model with initial mass 6.5M$_{\odot}$,
while it is log[X($^{17}$O)] $\sim -5.8$ for the 
3$M_{\odot}$ model.
The $^{18}$O abundance is dramatically decreased during the
second dredge-up, passing from log[X($^{18}$O)] $\sim -6$
to log[X($^{18}$O)] $\sim -10$. At the beginning of the
TPs phase some $^{18}$O is produced 
via proton capture by $^{17}$O,
so that its abundance rises up by 2 orders of magnitude
for the 6.5M$_{\odot}$ model (and by one order of magnitude
in the 3.5M$_{\odot}$ model). Like $^{17}$O, a maximum
value is reached, after which the surface $^{18}$O
abundance decreases as $^{16}$O is consumed within the 
envelope.

An important outcome of most of our models is that
the global C+N+O abundance of the ejecta is constant
within a factor of $\sim 2$. This can be seen in the
7th column of Table.~\ref{ejecta}, where we report the 
ratio between the global CNO abundance of the ejecta
and the initial value. With the only
exception of the 3M$_{\odot}$ model, for which the
effects of the 3rd dredge-up overwhelm those of HBB,
the values of the ratio between the average (C+N+O)
abundance of the ejecta and the initial value
are always $\leq 2$, being close to 1 for 
the most massive models.

These results are at odds with recent computations
of AGB models of the same metallicity by Fenner et al. 
(2004), where it was shown that:
\begin{itemize}
\item{Only models more massive than 6M$_{\odot}$ achieve
surface oxygen depletion.}
\item{The material expelled by AGBs is for all the 
computed masses both carbon and nitrogen rich.}
\item{The C+N+O strongly increases with respect to
the initial value.}
\end{itemize}
These findings led the authors to conclude that massive AGBs
may hardly have played a relevant role in the pollution of
the interstellar medium of GCs, since spectroscopic analysis
of NGC6752 stars found an anti correlation between [C/Fe] and
[N/Fe] independent of the luminosity (Grundahl et al. 2002);
besides the  C+N+O sum is approximately constant in many CGs
(Ivans et al. 1999).
The different convective model adopted is the reason for the
difference between our results and those obtained by Fenner 
et al. (2004) (see the
detailed discussion in paper I), the FST model
leading much more easily to efficient HBB which, in turn,
triggers larger luminosities, shorter AGB life-times, and a 
smaller number of 3rd dredge-up episodes.

\begin{figure}
\caption{The variation of the surface sodium abundance
       for the standard NACRE models.}
         \label{sodionacre}
\end{figure}

\subsection{Sodium nucleosynthesis}
Fig.~\ref{sodionacre} shows the behavior of surface 
sodium. In all cases we see an increase due to the 
second dredge-up, followed by an early phase of sodium 
production due to $^{22}$Ne burning during the first 
TPs. Later on, when the whole Ne-Na cycle is active, 
the sodium abundance declines. For 
M $\leq 5$M\sun\ the third dredge-up favors a 
later phase of sodium production, via proton capture
by $^{22}$Ne mixed into the envelope.
The $^{22}$Ne itself is a result of two 
$\alpha$ captures on the $^{14}$N mixed into the 
helium intershell at each third dredge up episode.

Particularly in the models with masses M$\leq 4$\msun\ 
a considerable amount of sodium is produced.
In the 3M$_{\odot}$ model, following each TP,
sodium is produced by $^{22}$Ne burning; during the
quiescent CNO burning phases the bottom of the envelope 
is hot enough to activate the Ne-Na chain, but not
to allow the Ne-Na reactions to act as a cycle (Arnould 
et al. 1999), which favors a large production of sodium.

The bottom panel of fig.~\ref{yield1} shows
the average sodium abundance of the ejected material,
as a function of the stellar initial mass. Sodium
is produced within the less massive models due to the
third dredge-up and to the modest sodium burning, but
is destroyed within the massive models, so that
the sodium content of the ejecta of these latter
is under abundant with respect to the initial value. 
We see that only the models with initial 
masses clustering around 4M$_{\odot}$ are able to 
expell material which is both sodium rich and 
oxygen poor, and so {\it only the abundances in
these ejecta would be in agreement with the
oxygen-sodium anti correlation} observed within GCs stars
(Gratton et al. 2001; Sneden et al. 2004).

\begin{figure}
\caption[f1.eps]{
Observed data which define the anticorrelation sodium vs. oxygen in the stars 
of several GCs. stars: M13, open squares: M3 (both from \cite{sneden2004}); 
full squares: NGC 6752 from \cite{grundahl2002}; full triangles: M4 and open 
triangles: M5 (both from \cite{ivans1999}); full dots: NGC 2808 from 
\cite{carretta2003}. Models by \cite{fenner2004} of 3.5, 5 and 6.5\msun, and 
models of this paper are also shown. The standard models
with $\eta=0.02$ (left line, points labelled by the mass
value) and the models with $\eta=0.10$ have been shifted by
0.3 dex in $[O/Fe]$ to be compared with the observations.
Equal masses are joined.
}\label{fenner}
\end{figure}

Also in regard to sodium, we note the different 
predictions of our models compared to those by 
Fenner et al. (2004), who expect extremely large 
sodium production for all the masses considered here.
(see their fig.1 and Fig.~\ref{fenner}). Within their models the
sodium produced is primary, and is produced via 
$^{22}$Ne burning, this latter being dredged-up 
from the inner helium layers.
In principle, this mechanism could work also in our
models (see fig. ~\ref{sodionacre}), but sodium 
production is made much less efficient due to:
{\it i)} the smaller number of 3rd dredge-up episodes;
{\it ii)} the larger temperatures, which favor sodium
destruction.
We therefore see that it is again the treatment of
convection the main reason of the differences
found in terms of the sodium content of the ejecta
of AGBs. It is interesting to note that, in terms
of the self-enrichment scenario, we have the opposite
problem compared to the Fenner et al. (2004) models:
they produce too much sodium, in great excess with
the increase observed in some GCs stars (which is
at most of $\sim 0.5$ dex), while in our case, for
the most massive models, we destroy it, as also
predicted by Denissenkov \& Weiss (2004).

Figure \ref{fenner} compares our results with 
those by Fenner et al. (2004) in the plane of oxygen 
versus sodium abundances, in which we have
reported several sets of observational data.
Our results should be shifted by +0.3dex in oxygen to 
be properly compared with the observations. We see 
that both sets are unable to reproduce the data.

\begin{figure*}
\caption{The variation of the magnesium and aluminum isotopes
       for the standard NACRE models of 3.5, 4 and 5\msun.}
         \label{mgalnacre}
\end{figure*}

\subsection{The magnesium and aluminum isotopes}

Figure \ref{mgalnacre} shows the variation of the magnesium and aluminum 
isotopes along the standard evolutions of 3.5, 4 and 5\msun. As the rate of 
proton capture on this isotope increases with the temperature at the bottom of 
the convective envelope, the $^{24}$Mg is more depleted for larger masses. 
Masses M$>5$\msun\ have qualitatively the same behavior, with more efficient $^{24}$Mg 
destruction. The heaviest destruction is found within the 6.5M$_{\odot}$ 
model, in which the surface final abundance is lower with respect to the 
initial value by a factor of $\sim 500$. We see from Tab.~\ref{ejecta} that the 
$^{24}$Mg abundance of the ejecta is lower the larger is the initial mass, 
reaching a minimum value of $[^{24}$Mg] $\sim -1.3$ for the 6.5M$_{\odot}$ 
model. The abundance of $^{25}$Mg, on the contrary, in a first stage 
increases due to the $^{24}$Mg proton capture during the first TPs, and later 
on its abundance decreases (e.g. in the 5\msun) due to burning to $^{26}$Al. 
The $^{26}$Al\ however decays into $^{26}$Mg only on a timescale of 7$\times 
10^5$yr, and so this element is a bottleneck for further proton capture on 
$^{26}$Mg, which leads to $^{27}$Al.  A direct path to $^{25}$Mg and $^{26}$Mg 
is through the third dredge up, as these isotopes are synthesized in the helium 
shell via capture of $\alpha$\ on $^{22}$Ne, and release respectively of a neutron 
or a gamma. This production mechanism is evident in Figure \ref{mgalnacre} for 
$^{26}$Mg, while it is also clear (especially in the left figure, relative to 
the 3.5\msun\ evolution) that the production of $^{25}$Mg is due to two 
mechanisms, dredge up and proton capture on $^{24}$Mg. As we do not have a 
large number of thermal pulses, the $^{27}$Al abundance can not rise by the 
huge factor (close to 10) shown by Globular Cluster stars (see Grundhal et al. 
2002, for the giants of the cluster NGC 6752). Further, the ratios between the 
magnesium isotopes are not consistent with the results by Yong et al. (2003), 
which indicate that $^{25}$Mg remains at 10\% of the $^{24}$Mg\ abundance, and 
$^{26}$Mg reaches at most $\sim 50$\%. The observational result both implies a 
not dramatic burning of $^{24}$Mg, and a mild, if any, increase in $^{25}$Mg 
and $^{26}$Mg. Notice that, in addition, we have to count into the $^{26}$Mg 
abundance also the abundance of the unstable isotope $^{26}$Al, and the result 
is at variance with observations. 

In spite of the not good agreement of these abundances with the observation, at 
least the trend of our models is in the right direction, as the ratios 
$^{25}$Mg/$^{24}$Mg and $^{26}$Mg/$^{24}$Mg do not exceed $\sim 3$\ for masses 
up to 5\msun. The corresponding models by Fenner et al. (2004), in which the 
smaller efficiency of convection allows a longer evolutionary phase and many 
episodes of third dredge up provide ratios larger than 100. Notice also that 
for elements whose abundances are very small, also the initial abundances and 
the exact modeling of the thermal pulses may influence strongly the results. 
The central part of Figure \ref{mgalnacre} in fact shows that the evolution of 
the 4\msun\ suffers an anomalous episode of third dredge up, which we are 
uncertain whether to attribute to numerics or to a real effect. This lonely 
episode changes the surface abundances of sodium and magnesium in such a way 
that the resulting yields of the elements having low abundances are affected, 
although the most abundant yields (e.g. CNO) are not. This requires an 
additional detailed study before we can reject or accept these results as 
conclusive for the problem of abundance variations in GCs.  


\subsection{The lithium content of the ejecta}
We conclude this general description with lithium, which
is created during the first TPs via the Cameron \& Fowler
(1971) mechanism, and then destroyed as soon as $^3He$ 
is extinguished in the envelope. 
We see from the 3rd column of tab.~\ref{ejecta} that
the lithium content of the ejecta first decreases
with increasing mass. In fact, the larger 
is the mass, the hotter
is the base of the convective envelope, the more
rapidly $^3$He is destroyed, the shorter is the 
phase during which the star shows up as lithium-rich;
for models more massive than 5M$_{\odot}$, as
already discussed for the oxygen content of the ejecta,
we have that the mass loss is so strong during the 
first TPs that a considerable fraction of the mass
is lost when lithium has been produced and not yet
destroyed. The differences among the lithium
abundances of the various models is however within
a factor of $\sim 2$, and is about a factor 2
smaller than the average
abundance which is observed in population II stars.

\begin{figure}
\caption{The average chemical content of the ejecta
      of the standard NACRE models as a function
      of the initial mass. {\bf Top}: Oxygen 
      abundance; {\bf Bottom}: Sodium abundance.}
         \label{yield1}
\end{figure}

\subsection{The overall chemistry of the ejecta}
In examining the overall chemical content of the ejecta
of our models, there are four common features, which
hold independently of mass:
\begin{enumerate}
\item{
The lithium content is within a factor 2
of the value observed in population II stars.}

\item{The ejecta are helium rich: particularly
models with initial mass $M>4M_{\odot}$
during the second dredge-up reach surface 
helium mass fractions of $Y\sim 0.32$. Helium
is also produced during the AGB phase during any
interpulse phase, but the consequent overall
increase of $Y$ is limited in all cases to
$\delta Y\sim 0.005$.}

\item{When compared with the initial abundances, we see that the material lost 
during the evolutions is enhanced in $^{14}N$ by at least a factor of $\sim 10$ 
and depleted in  $^{12}C$ by a factor of $\sim 4$. This trend is consistent 
with the observations, indicating CN cycled composition and no carbon 
enhancement (Cohen et al. 2002, Cohen \& Melendez 2004). However, the
anticorrelation between carbon and nitrogen is not found in the yields 
as a function of the initial mass (see Table 2).}

\item{The (C+N+O) abundance is constant within a
factor of $\sim 2$, due to the small number of
3rd dredge-up episodes.} 
\end{enumerate}

For the other elements, the stellar yields depend
sensibly on the stellar mass, both quantitatively and
qualitatively.
The more massive models pollute the interstellar
medium with material which is oxygen and (partially)
sodium depleted. $^{24}$Mg is heavily depleted 
(a factor of $\sim 10$ with respect to the initial
value), so that high magnesium isotopic ratios
(larger than solar) are expected.

Conversely, within models less massive than 
$\sim 4.5M_{\odot}$, the temperature at the base of the 
external zone is such that the Ne-Na cycle is only 
partially activated, and only in the last AGB
phases; consequently, the ejecta of these stars
are sodium rich and oxygen poor. The $^{24}$Mg
depletion is negligible due to the low temperatures,
therefore the magnesium isotopic ratios are reduced,
although not at the level which would provide 
agreement with the relevant observations by 
Yong et al. (2003). Further, the magnesium - aluminum 
anti correlation
(see e.g. Grundahl et al. 2002) is not
fully reproduced, as the $^{27}$Al production is not 
very efficient.

\section{NACRE vs CF88}
The nuclear cross-sections play a delicate role in
determining the main physical and chemical properties
of the AGB evolutions.
From a physical point of view, a variation of the 
cross-sections of the nuclear reactions mostly 
contributing to the global energy 
release might influence the thermal stratification 
of the star; from a chemical
point of view, a change in the cross-sections of 
those reactions involving key elements like sodium or
magnesium, though energetically not relevant,
might alter the equilibrium abundances, hence 
the average chemical content of the ejecta.

We explore the uncertainty of
the results connected with the cross-sections of the 
various reactions included within our network, by
performing a detailed comparison
between the results presented in the previous section
and those obtained with the CF88 release, which are
still widely used in modern AGB computations. 
This was made also to have an idea of 
the degree of uncertainty of the results 
connected with the cross-sections of the 
various reactions included within our network.
We calculated a new set of models with the same 
physical and chemical inputs of the NACRE models, 
but adopting the CF88 rates for the nuclear 
cross-sections. 

\begin{figure}
\caption{The comparison between the time-scale
     for helium burning ({\bf Top}) and the 
     ratio between the time-scales for helium
     and for hydrogen burning ({\bf Bottom})
     for the NACRE (solid) and CF88 (dotted)
     models of intermediate mass.}
         \label{timevol}
\end{figure}

We didn't find any appreciable difference during the 
MS evolution for the whole range of masses involved,
as the cross-sections of the relevant reactions
are the same for both sets of models. For each value
of the initial mass, we could verify that the duration
of the H-burning phase, the innermost
point reached during the first dredge-up and 
the consequent changes in the surface chemistry
are unchanged.

The first differences between the models 
appear in the duration of the helium burning phase,
as can be seen in fig.~\ref{timevol}. 
Also the ratio between the times of helium
and hydrogen burning are consequently affected.
The reason of this difference stands 
in the rate of the reaction 
$^{12}$C+$\alpha \rightarrow ^{16}$O, which 
is larger by a factor of $\sim 1.7$
in the NACRE case. This leads to slightly 
longer time-scales for helium burning (see 
the discussion in Imbriani et 
al. 2001 and Ventura \& Castellani 2005).

The AGB evolution of the models is physically
very similar, because the global nuclear energy 
release during the quiescent phase of CNO burning 
(which, we recall, is for most of the time the 
only nuclear source active within the star) is
dominated by the proton captures by $^{12}$C, 
$^{13}$C and $^{14}$N nuclei, whose corresponding 
cross-sections are similar in the two cases.
We could verify that the duration of the whole
AGB phase for the two sets of models, as well as
the temporal evolution of the most relevant
physical quantities, are essentially the same.

\begin{figure}
\caption{The average chemical content of the 
   ejecta of the NACRE (solid) and CF88
   (dotted) models of intermediate mass in
   terms of lithium, carbon, nitrogen and
   oxygen abundances.}
         \label{conyie1}
\end{figure}

In the four panels of fig.~\ref{conyie1} 
we show the average chemical content of the 
ejecta, in terms of lithium and CNO abundances. 
We see that the lithium content is extremely 
similar for the two sets of models, 
while the $^{12}$C, $^{14}$N and $^{16}$O 
abundances are lower in the CF88 models, 
when compared to NACRE. 
This difference is due to the cross-sections 
of the reactions of proton capture by $^{17}$O 
atoms, which, as already discussed in Sect.2,
are much lower in the CF88 case.
The three left-panels of fig.~\ref{sezurto} show, 
respectively, the variation with temperature of 
the ratio (CF88/NACRE) between the rates of the 
reactions $^{17}$O(p,$\gamma)^{18}$F (top panel),
$^{17}$O(p,$\alpha)^{14}$N (middle panel),
and of the ratio 
$\sigma(^{17}$O(p,$\gamma)^{18}$F)/$\sigma(^{17}$O(p,$\alpha)^{14}$N)
in the two cases. We can see that in the range of temperatures
of interest here ($7.5\leq \log T \leq 8$) the reaction
$^{17}$O(p,$\alpha)^{14}$N is more efficient in the NACRE case 
by a factor of a few hundred, while the difference for
the reaction $^{17}$O(p,$\gamma)^{18}$F is a factor
of $\sim 5$. In the third panel, more important, we can see
that the favorite channel of $^{17}$O destruction switches
from $^{17}$O(p,$\alpha)^{14}$N to $^{17}$O(p,$\gamma)^{18}$F
passing from the NACRE to the CF88 cross-sections.

\begin{figure*}
\centering{
           }

\caption{
   {\bf Left}: Variation with temperature of the
   logarithm of the ratio between the CF88 and the NACRE
   cross section of the reaction $^{17}$O(p,$\gamma)^{18}$F
   ({\bf top}), $^{17}$O(p,$\alpha)^{14}$N 
   ({\bf middle}) and of the
   ratio $\sigma(^{17}$O(p,$\gamma)^{18}$F)/
   $\sigma(^{17}$O(p,$\alpha)^{14}$N) ({\bf bottom}). 
   {\bf Right}: The same as the left panel, but for 
   reactions $^{23}$Na(p,$\gamma)^{24}$Mg and 
   $^{23}$Na(p,$\alpha)^{20}$Ne.}
      \label{sezurto}
\end{figure*}

This, in turn, has two important consequences:
\begin{itemize}

\item{In the CF88 models we have a much larger
production of the heaviest oxygen isotopes.}

\item{In the NACRE case, the equilibrium abundance
of $^{17}$O is much lower, and the nucleosynthesis
favors a return to $^{14}$N rather than a production
of $^{18}$F: this explains the differences between
the CNO abundances of the ejecta of the two sets of models
which can be seen in fig.~\ref{conyie1}.}

\end{itemize}

\begin{figure}
\caption{Average chemical content of the ejecta
       of IMS models calculated with NACRE (solid)
       and CF88 (dotted) sets of nuclear 
       cross-sections in terms of $^{23}$Na and
       $^{24}$Mg. The bottom panel show the
       isotopic ratios of magnesium.}
         \label{conyie2}
\end{figure}

\noindent
Turning to heavier elements, we show in the 
four panels of fig.~\ref{conyie2} 
the abundances of $^{23}$Na
and $^{24}$Mg of the ejecta of the different models,
and the ratio of the magnesium isotopes,
$^{25}$Mg/$^{24}$Mg and $^{26}$Mg/$^{24}$Mg. We note
from the left-upper panel of fig.~\ref{conyie2} that sodium
production can be achieved efficiently in the NACRE
case, while it is not present in the CF88 models.
This can be explained very simply on the basis
of the cross-section of the reaction 
$^{22}$Ne(p,$\gamma)^{23}$Na, which, in the range
of temperatures relevant in this case, is lower
in the CF88 case by at least a factor of $\sim 100$,
reaching a maximum difference of a factor 
of $\sim 2000$ for $\log(T)=7.8$. 

\begin{figure}
\caption{The AGB evolution of the surface abundances
       of $^{23}$Na ({\bf top}) and $^{22}$Ne 
       ({\bf bottom}) of two models with initial 
       mass 5M$_{\odot}$ calculated with the NACRE 
       (solid) and CF88(dotted) nuclear cross-sections.}
       \label{nane5}
\end{figure}

In the two panels of fig.~\ref{nane5} we compare the variation 
with mass of the $^{22}$Ne and $^{23}$Na surface
abundances within two models of initial mass
5M$_{\odot}$ calculated with the NACRE and CF88
nuclear cross sections.
We see an early phase of sodium production and neon
destruction during the first thermal pulses in the 
NACRE model, and a later increase of the sodium
abundance due the dredging-up of $^{22}$Ne, which
is later converted to $^{23}$Na. In the CF88
case the surface $^{22}$Ne never decreases,
and sodium is destroyed when the temperatures at the
base of the external zone become large enough to
activate efficiently the Ne-Na cycle.

The top and middle right panels of fig.~\ref{sezurto} 
show the ratio of the cross-sections corresponding 
to the two channels of sodium destruction 
($^{23}$Na(p,$\gamma)^{24}$Mg and 
$^{23}$Na(p,$\alpha)^{20}$Ne) (always in terms of 
CF88/NACRE value) and the ratio
$\sigma(^{23}$Na(p,$\gamma)^{24}$Mg)/$\sigma(^{23}$Na(p,$\alpha)^{20}$Ne)
in the two cases. In the NACRE case 
sodium is destroyed more easily, and the favorite 
channel is magnesium production in the relevant 
range of temperatures; this determines a larger 
$^{24}$Mg equilibrium abundance, and explains the 
difference between the models which can be seen 
in the right-upper panel of fig.~\ref{conyie2}.

Turning to the magnesium isotopic ratios, 
$^{25}$Mg/$^{24}$Mg is similar in the 
two cases (see the left-lower 
panel of fig.~\ref{conyie2}), while $^{26}$Mg/$^{24}$Mg 
is lower in the CF88 models; the reason is that 
the rate of the reaction $^{26}$Mg(p,$\gamma)^{27}$Al 
is a factor of $\sim 10$ larger in the CF88 case, thus 
favouring $^{26}$Mg destruction in favor of $^{27}$Al 
production.

By comparing the AGB models calculated with
the two sets of cross-sections we conclude that the physical
behavior is essentially the same, because the rates
of the reactions mostly contributing to the global energy
release are scarcely changed. In terms of nucleosynthesis
(and therefore of the average chemical content of the
ejecta) we find important variations only for sodium and
the heavier isotopes of oxygen. The former is not produced
at all in the CF88 models (contrary to the NACRE case) 
due to the extremely low cross-section of the $^{22}$Ne 
proton capture reaction; the equilibrium abundances of
$^{17}$O and $^{18}$O are much lower in the NACRE case,
because of the larger values of both the $^{17}$O proton
capture reactions.

\section{The role of mass loss}
The effects of mass loss on the evolution of AGB stars
is well documented in the literature (Sch\"onberner 1979): 
mass loss halts the increase of luminosity, 
and progressively peels off all the envelope, 
leaving eventually a central remnant
which evolves as a white dwarfs. It determines a 
general cooling of the structure, therefore reducing
the intensity of HBB at the base of the external convective
zone. Since the effects of mass loss 
become evident only when the mass of the envelope 
is considerably reduced, models
with different mass loss rates will differ only in the 
terminal part of their evolution, while the general
physical behavior during the first TPs is unchanged.

The NACRE and CF88 models presented in the previous sections
were calculated with the parameter $\eta_R=0.02$, 
in the Bl\"ocker's formula.
This choice is due to a previous calibration, made on 
the basis of a detailed comparison
between the observed and the theoretical luminosity
function of lithium rich AGB stars in the Magellanic
Clouds (Ventura et al. 2000). For our models, a value of 
$0.01 \leq \eta_R \leq 0.02$ for stars with initial 
mass in the range 3 $\leq$ M/\msun $\leq$ 4.5 is able to 
reproduce the observed trend of surface lithium vs
luminosity which is observed in the Clouds.

We cannot completely rule out the possibility that 
the parameter $\eta_R$ to be used during the AGB 
evolution might show a dependency on the metallicity 
(we recall that the models discussed 
in this paper have a metallicity which is a factor of 
10 lower than the LMC stars) and on the stellar mass,
or that the mass loss is heavily influenced by the
environment.
In order to test the level of uncertainty which is
connected with the mass loss, we decided to explore 
the sensitivity of our results on changes in the
value of $\eta_R$, and we discuss it for 
for two representative examples of our stars. 

\begin{figure}
\caption{Variation with time of the total mass
        of three models of initial mass $5M_{\odot}$
        calculated with three different value of the
        parameter for mass loss, $\eta_R$. Time
        has been set to 0 at the beginning of the AGB
        evolution.}
         \label{conmassa5}
\end{figure}

\subsection{Massive AGBs}
We compare the standard model 
of initial mass 5M$_{\odot}$ already presented in 
Sect. 3 (eta002 model) with two models of the 
same initial mass, computed by assuming, 
respectively, $\eta_R=0.1$ (eta010 model) and 
$\eta_R=0.2$ (eta020 model). Fig.~\ref{conmassa5} 
shows the variation with time of the mass.
Times have been set to 0 at the beginning of the 
AGB evolution. We see that the total duration of 
the AGB phase is strongly dependent on $\eta_R$, 
ranging from $t_{AGB} \sim 73,000$ yr for the eta002 
model down to $\sim 27,000$ yr in the eta020 case.
The eta002 model reaches a maximum luminosity of
$66,000 L_{\odot}$ during the 20th interpulse period,
while the eta010 and eta020 models achieve a
maximum luminosity of, respectively, $50,000 L_{\odot}$
(13th interpulse period), and $43,500 L_{\odot}$
(10th interpulse period). 
In conjunction with the maximum luminosity, all the models
also attain the largest temperature at the base of the
external envelope. The maximum $T_{bce}$ are:
$T_{bce}=103 \times 10^6$K (eta002), 
$T_{bce}=100 \times 10^6$K (eta010) and 
$T_{bce}=98 \times 10^6$K (eta020). These values are
quite similar, especially when compared to the differences in the
maximum values of luminosity reached by the three models.
This is not surprising, as a main
feature of the FST convective model is that, within
the most massive AGBs, it leads to a very efficient 
HBB already during the first TPs: even a strong
increase of mass loss cannot prevent the base
of the external zone to become extremely hot. On the
basis of these results, we may expect a deep nucleosynthesis
to take place at the base of the external convective zone
even in the eta020 model.

\begin{figure*}
\centering{
           }
\centering{
           }

\caption{Variation with time and total mass of the
         CNO surface abundances of the same models
         presented in fig.~\ref{conmassa5}. The
         right-lower panel shows the total C+N+O
         abundance. The abundances are given
         in mass fraction.}
   \label{confcno5}
\end{figure*}

In the four panels of fig.~\ref{confcno5} 
we show the evolution of the
surface CNO abundances for the three models, plus the
variation of the total C+N+O abundance. For each of these
elements we show both the variation with time and with mass.
In the left-upper panel we show the variation of surface
$^{12}$C. In the top of this panel we see that
the temporal evolution is very similar, with an early 
phase of destruction at the beginning of the AGB evolution
followed by a later phase of production, when the temperatures
at the base of the external zone are sufficient to allow
the full CNO cycle to be activated, and the effects
of the 3rd dredge-up become more evident. 
The only difference among the three models
is that the AGB evolution is halted earlier for larger
values of $\eta_R$. Since for
all the models the evolution stops when the carbon
abundance was increasing, this acts in favor of
a larger $^{12}$C content of the ejecta for lower
mass loss rates. In the lower part of this panel we see
the evolution of $^{12}$C with 
the stellar mass. The 
above effect is partly compensated by the fact that,
for larger $\eta_R$, the star looses a not negligible 
fraction of its mass when the carbon abundance was
still unchanged, even before the early phase of
destruction at the beginning of AGB. This is the reason
why the average $^{12}$C abundance of the ejecta of our
models show a maximum difference of $\sim 0.1$ dex,
and is therefore consistent with the value
$[^{12}$C]=--0.7 given in Sect. 3.

An analogous discussion can be made for nitrogen, as can 
be seen in the right-upper panel of fig.~\ref{confcno5}. 
The surface $^{14}$N increases in all cases, because
nitrogen is created at the base of the external envelope
due to HBB and, in the final part of the evolution, also
due to the effects of the 3rd dredge-up. Again we note a strong
similarity in the temporal evolution, the only difference
being that in the large $\eta_R$ models the $^{14}$N
content of the ejecta is expected to be lower because
the AGB evolution is halted earlier. In reality, at odds
with the $^{12}$C case, we expect a larger nitrogen
content of the ejecta of the eta002 model because
this latter case loses less mass at the very beginning of
the AGB evolution, when the $^{14}$N abundance was still
unchanged since the second dredge-up (see
the lower part of the right-upper panel 
of fig.~\ref{confcno5}). 
The $^{14}$N average content of the ejecta is 
therefore more dependent on mass loss, ranging 
from $[^{14}$N]=1.15 in the eta002 model, down to 
$[^{14}$N]=0.83 in the eta020 model; the eta010 model, 
with $[^{14}$N]=0.92, shows an intermediate behavior.
The global spread of the $[^{14}$N] value varies
at most by a factor of $\sim 2$ if $\eta_R$ varies
by one order of magnitude.

As already pointed out, the temperatures at the base
of the convective zone are sufficiently large to
activate the full CNO cycle in all the models, so that
in all cases we have a certain amount of $^{16}$O depletion,
as can be seen in the left-lower panel of fig.~\ref{confcno5}.
The trend of the $^{16}$O average content of the ejected
material for various $\eta_R$ is straightforward:

\begin{itemize}

\item{For larger values of $\eta_R$ a consistent part
of the envelope mass ($\Delta M \sim 1M_{\odot}$) is
lost when the oxygen abundance is still unchanged.}

\item{Since the temperature at the base of the envelope
reached by the large $\eta_R$'s models is lower,
in these cases $^{16}$O is destroyed less heavily,
and this leads to higher oxygen equilibrium abundances.}

\end{itemize}

We thus find that a strong oxygen destruction is 
hardly found within the models with the largest 
mass loss rates. The $[^{16}$O] of the ejecta 
is more dependent than $[^{14}$N] on the 
assumed $\eta_R$: we have $[^{16}$O]=--0.6 
for the eta002 model, $[^{16}$O]=--0.3 in the 
$\eta_R=0.1$ case, and $[^{16}$O]=--0.15 for 
$\eta_R=0.2$. 

In the right-bottom panel of Fig.~\ref{confcno5} 
we show the
total C+N+O abundance. In the eta010 and eta020 
models the sum of the CNO abundances is constant,
because mass is lost so rapidly that the effects of the
3rd dredge-up are negligible. In the eta002 model, 
during the last TPs, carbon is efficiently dredged-up, 
and is later converted to $^{14}$N by HBB; however, the 
total increase of the C+N+O is within $\sim 0.2$ dex. 
We may therefore conclude that within the FST framework 
the most massive AGB models show surface C+N+O abundances 
which are constant within a factor of $\sim 2$, 
independently of mass loss.
 
\begin{figure}
\caption{AGB evolution of the surface abundances of
     $^{23}$Na (top) and $^{22}$Ne (bottom) of the
     same models presented in fig.~\ref{conmassa5}.}
     \label{confna5}
\end{figure}

Turning to heavier nuclei, we focus our attention
on sodium. Fig.~\ref{confna5} shows the 
variation of the surface sodium 
abundance as a function of time (top panel) and
mass (bottom). An early phase of production, due to proton
capture by $^{22}$Ne nuclei, is followed by a phase
of sodium destruction when the temperatures at the
base of the outer convective zone become large
enough that sodium is destroyed by proton capture.
A larger mass loss rate acts in favor of larger 
sodium yield because a large fraction of the stellar 
mass is lost
when sodium is produced, and also because the evolution
is halted when the surface sodium has not yet 
been completely destroyed (bottom panel).

Actually, sodium turns out to be the element 
most sensitive to variations of the mass loss 
rate. A larger $\eta_R$ changes completely 
the situation, in the sense that now we expect 
the mass expelled by massive AGBs to be sodium 
rich (with respect to the initial mass fraction)
rather than sodium poor. $[^{23}$Na]
linearly increases with $\eta_R$, while a 
positive sodium yield is not possible at $5M_{\odot}$
with the standard $\eta_R=0.02$ value. The oxygen 
yield shows a similar
behavior, though in this case the convection is so
efficient that $[^{16}$O] is negative in all
cases. A simultaneous sodium production and oxygen
depletion in the chemistry of the ejecta, in agreement
with the observed anti correlation, is possible
only for $\eta_R \sim 0.1$.

Turning to magnesium, we find that $^{24}$Mg is depleted
in all cases, but the final abundance is a factor of
$\sim 500$ lower in the eta002 case, while it is just a
factor of 2 lower in the eta020 model. The $^{24}$Mg
abundance of the ejecta is a factor of $\sim 10$
lower than the initial value for the eta002 model,
while it is lower by only $\sim 0.1$ dex for
$\eta_R=0.20$. 

\noindent
The average content of $^{25}$Mg is not strongly dependent
on the mass loss rate, because it reaches a maximum value
and then declines as $^{24}$Mg is destroyed; in reality, 
in the eta010 and eta020 models the evolution is completed
when the $^{25}$Mg is almost at its maximum value. The net
result is that within $\sim 0.1$ dex we find 
$[^{25}$Mg]=0.5 for all the models.
The situation is different for the heaviest isotope,
because the surface $^{26}$Mg increases for the whole
evolution. In this case a larger mass loss rate leads
to a lower final abundance, so that the yield is lower.
For the eta002 model we find $[^{26}$Mg]=0.35, while
$[^{26}$Mg]=0.15 in the $\eta_R=0.01$ case. The average
$^{26}$Mg content of the eta020 model is practically 
unchanged with respect to the initial value.

In terms of isotopic ratios, lower values of both
$^{25}$Mg/$^{24}$Mg and $^{26}$Mg/$^{24}$Mg are expected
for larger mass loss rates, because in that case we
have a lower depletion of $^{24}$Mg. In agreement
with that, we find isotopic ratios $\sim 3$ for
$\eta_R=0.02$, $\sim 1$ for $\eta_R=0.1$ and
$\sim .4$ for $\eta_R=0.2$, the $^{25}$Mg/$^{24}$Mg
ratio being always slightly larger than the 
$^{26}$Mg/$^{24}$Mg. 

\subsection{The 4M$_{\odot}$ model.}
The situation is a bit more complex for 
lower masses, because in that
case the temperature reached by the base of
the outer convective zone never exceeds
$\sim 10^{8}$ K, therefore they achieve only
in a later phase of their AGB evolution the
conditions which are necessary to trigger a
deep nucleosynthesis within the convective
envelope. In these cases, at odds with the
most massive models, we expect that a stronger
mass loss, triggering an earlier cooling of
the structure, may prevent some reactions to 
occur at all. 

We therefore calculated a model with initial
mass M=4M$_{\odot}$ with a parameter for mass
loss $\eta_R=0.10$ (eta010 model), and we
compare it with the model with the same initial
mass calculated with $\eta_R=0.02$, presented
in Sect. 3.
The total duration of the AGB phase for the eta010 model
is shorter, as expected: The total mass of the star
reduces to $\sim 1.4M_{\odot}$ within 
$t_{AGB}\sim 87,000$ yr, to be compared to 
$t_{AGB}\sim 150,000$ yr of the eta002 model.

\begin{figure}
\caption{The evolution of luminosity (top) and temperature
    at the base of the envelope (bottom) of two models
    with initial mass 4M$_{\odot}$ calculated with two
    values of the parameter entering Bl\"ocker (1995)
    prescription for mass loss: $\eta_R=0.02$ (solid track)
    and $\eta_R=0.1$ (dotted).}
     \label{confphys4}
\end{figure}

\begin{figure}
\caption{Comparison between the depletion of
    surface oxygen within two models with initial mass
    4M$_{\odot}$ calculated with two different values
    of $\eta_R$.}
     \label{confo164}
\end{figure}

In fig.~\ref{confphys4} we compare 
the variation with time of the
luminosity and of the temperature at the base of the
envelope of the two models, as a function of the
AGB time. We see from the top panel that there is
a difference of $\sim 0.2$ dex between the maximum
value of the luminosity reached by the two models, while,
in terms of temperature, the base of the convective zone
of the eta002 model achieves a maximum value of
$T_{\rm bce}=95\times 10^6$ K, to be compared
to the maximum temperature $T_{\rm bce}=88\times 10^6$ K 
reached in the $\eta_R=0.10$ case.
In terms of the chemical content of the ejecta,
we may repeat for $^{12}$C and $^{14}$N the same 
discussion performed for the 5M$_{\odot}$ model,
because the temperatures in this case, though
lower, are still sufficient to favor an early
phase of $^{12}$C destruction followed by a later
phase of production, and a progressive increase
of the surface $^{14}$N abundance due both to
HBB and to the effects of the 3rd dredge-up.
Thus, the $^{12}$C abundance
of the ejecta is almost the same for both 
models, while the $^{14}$N abundance is lower
in the eta010 model by a factor of $\sim 2$.

The different values of the temperatures reached
at the base of the outer convective zone in the
two models lead to a different degree of the
oxygen depletion at the base of the envelope, as
can be seen in the two panels of fig.~\ref{confo164}, 
where we show the variation of the surface $^{16}$O
with time (top panel) and mass (bottom).
We see that both models start to deplete oxygen
after $\sim 40,000$ yr, but the depletion is made
difficult in the eta010 models by the lower
temperatures, so that the final abundance is only
$\sim 0.15$ dex lower than the $^{16}$O present in
the envelope at the beginning of the TPs phase.
In the eta002 model a stronger depletion is achieved.
The average oxygen content of the ejecta is
$[^{16}$O]=--0.4 in the eta002 model, while it
is $[^{16}$O]=--0.1 for $\eta_R=0.1$.
Even for the $M=4M_{\odot}$ model we find that
the C+N+O abundance is constant within a factor 
of $\sim 2$ for the whole evolution, the eta002
models showing the largest increase due to
the higher number of 3rd dredge-up episodes.

The situation concerning sodium is more tricky.
The eta002 model, after the initial phase of
production, destroys sodium more efficiently
due to the larger temperatures reached; yet, the
evolution is so long that some sodium is dredged-up 
later on; as a consequence, the sodium content
of the ejecta is almost the same in the two cases,
i.e. $[^{23}$Na]=0.5. The less massive models are
therefore efficient sodium producers, independently
of the mass loss rate adopted.
In terms of the oxygen-sodium anti correlation, the
eta002 model in this case is consistent with a
simultaneous oxygen depletion and sodium production,
while only a poor oxygen depletion is expected
for larger mass loss rates.

As for magnesium, the situation is deeply different
with respect to the 5M$_{\odot}$ case. The temperatures
here are not sufficiently high to favor an efficient
magnesium destruction, so that even in the eta002
model $^{24}$Mg is reduced by only a factor of $\sim 2$.
For $\eta_R=0.1$ the surface $^{24}$Mg is almost 
unchanged.
Even for the heavier isotopes the production is
much lower than in the 5M$_{\odot}$ model. 
In terms of the isotopic ratios, we find 
$^{25}$Mg/$^{24}$Mg $\sim 0.9$ and 0.2, respectively,
for $\eta_R=0.02$ and $\eta_R=0.1$, while
$^{26}$Mg/$^{24}$Mg is 0.7 for $\eta_R=0.02$ and
0.2 for $\eta_R=0.1$.

We may therefore summarize the influence of mass
loss on the AGB models as follows:

\begin{enumerate}

\item{The carbon content of the ejecta is almost 
independent of mass loss, while the nitrogen abundance
may vary by a factor of $\sim 2$, lower $\eta_R$
models showing the larger enhancement. This holds
for all the masses calculated, because at least the
CN cycle is always operating.}

\item{The oxygen abundance of the ejecta proves to be
more sensitive to mass loss; a lower $\eta_R$ favors 
larger oxygen depletion. For all the masses considered
we achieve oxygen depletion for $\eta_R=0.02$, while 
a poor depletion is expected for larger values of
$\eta_R$, particularly for the lowest masses.}

\item{The C+N+O sum is in all cases constant within
a factor of $\sim 2$, independently of mass and mass 
loss.}

\item{Larger $\eta_R$ favor sodium production in the more
massive models, because in that case the AGB evolution is
halted when the surface sodium has not yet been destroyed.
The less massive models are efficient sodium producers, 
independently of the mass loss rate.}

\item{The isotopic magnesium isotopes keep below unity
independently of mass loss for the less massive models.
For larger masses we have a steeper dependence on $\eta_R$:
we find $^{25}$Mg/$^{24}$Mg $\sim^{26}$Mg/$^{24}$Mg $\sim 3$
for $\eta_R=0.02$, down to 
$^{25}$Mg/$^{24}$Mg $\sim^{26}$Mg/$^{24}$Mg $\sim 0.4$ for 
$\eta_R=0.2$.}

\end{enumerate}

\section{Which implications for the self-pollution
scenario?}

There is still a strong debate concerning the role 
which AGBs may have played in the pollution of the 
interstellar medium of GCs: Fig.~\ref{fenner} 
shows that we are far from being able to falsify the hypothesis
that the chemical content of their ejecta
may account for the chemical anomalies observed
in GCs stars (Denissenkov \& Herwig 2003; 
Fenner et al. 2004; Ventura et al. 2002; paper I).

Within the MLT framework for the treatment of
convection the most recent work by Fenner et al.
(2004) shows that it is hardly possible to reconcile
the theoretical findings with the observational
scenario, because the expected chemical content
of the ejecta show a largely increased value
of the global C+N+O abundance, a very poor oxygen
depletion, and an extremely large sodium production.
These results are all in contrast with the 
observational evidence.
Their findings were confirmed by our AGB models
calculated with the MLT convection, presented
and extensively discussed in paper I.
 
If the FST model is used, due to the larger
temperatures reached at the bottom of the external
envelope, we find on the contrary that the C+N+O
is always constant within a factor of $\sim 2$,
in agreement with the results of Ivans et al. (1999).
Also, oxygen depletion is easily achieved in all
models more massive than $3M_{\odot}$.  

The FST models in the range $3.5 - 4.5$M$_{\odot}$
with a mass loss rate in agreement with the 
calibration given in Ventura et al. (2000)
pollute the interstellar medium with material
having a chemistry qualitatively in agreement with the
chemical anomalies observed, that is:

\begin{itemize}

\item{The C+N+O sum is almost constant}

\item{Oxygen and sodium are anticorrelated,
in qualitative agreement with the observed trend.
From a quantitative point of view a larger sodium
content, coupled to an even stronger oxygen depletion
would be necessary to match the observational
evidence.}

\item{The magnesium isotopic ratios are well below 
unity, but the $^{26}$Mg/$^{24}$Mg
ratio is larger than the $^{25}$Mg/$^{24}$Mg 
contrary to observations (Yong et al. 2003). In addition, 
the magnesium vs. aluminum anti correlation is not reproduced.}

\end{itemize}

In the most massive models, on the contrary, the 
temperatures are so large that sodium, after an 
early phase of production at the beginning of the 
AGB phase, is destroyed; with the standard 
$\eta_R=0.02$ value for the mass loss rate 
parameter we expect to have a negative [Na/Fe]
and a correlation between sodium and oxygen.
One further problem in this case would be the
extremely high values of the magnesium isotopic
ratios, which cluster around $\sim 3$, duo to 
strong $^{24}$Mg burning. If a larger
mass loss rate is adopted, we expect a poor oxygen 
depletion and sodium destruction, which would be 
more consistent with the observations. The
magnesium isotopic ratios would be in this case
below unity, which is also in better agreement
with the observed abundances (Yong et al. 2003).  

The CF88 models share almost the same properties
of the NACRE models, with the only difference of
sodium, which can not be produced in this
latter case, because of the extremely low values
of the $^{22}$Ne proton capture reaction
cross-sections.
 
\section{Conclusions}
We present AGB models of intermediate mass in the
range $3M_{\odot} \leq M \leq 6.5M_{\odot}$ with
metallicity Z=0.001. 

This work, together with the results of paper I,
explores the role of some parameters of AGB evolution and helps to
understand that it is affected by so many uncertainties
that it is still implausible to use the results of a unique set of
AGB computations to falsify the self--enrichment scenario for
globular cluster stars. Nevertheless, these results show trends
in the yields which are different from those of other researchers, 
and may help to find a way for the solution of the self--enrichment
problem.
In particular,
contrary to several recent AGB computations, and thanks to
the use of the FST model, we find that convection 
at the base of the external  
zone during the quiescent phase of CNO burning is 
so efficient to lead to extremely high temperatures
($T_{\rm bce} \sim 10^8$K), sufficient to trigger strong
HBB. We find that oxygen is depleted in all cases
with the only exception of the 3M$_{\odot}$ model.

Our main findings are the following:
\begin{enumerate}
\item{
The physical behavior of the models turns out to be
independent on the nuclear cross-sections used, the
results obtained with the NACRE rates being very similar
to CF88: this is due to the fact that the reactions
most contributing to the global energy release, i.e.
proton captures by carbon and nitrogen nuclei, are
the same in the two cases.}

\item{
One strong prediction, which holds independently of
mass, is that the total C+N+O abundance of the
ejecta is almost constant, ad odds with previous 
investigations.}

\item{
Sodium is produced within the NACRE models with 
M $< 4.5$\msun, while it is destroyed in 
the more massive stars, due to an efficient 
action of the Ne-Na cycle. In the CF88 models
sodium is systematically destroyed, because
of the extremely low values of the $^{22}$Ne proton 
capture reaction.}

\item{
In terms of magnesium isotopes, the 
$^{25}$Mg/$^{24}$Mg and $^{26}$Mg/$^{24}$Mg ratios 
are well below unity in the models with 
M $<$ 4.5M$_{\odot}$, while they reach values 
approaching $\sim 10$ for M$>$5M$_{\odot}$, 
due to strong $^{24}$Mg burning.}

\item{
Mass loss influences the global duration of the AGB
life; it also determines a general cooling of the 
structure before some nuclear reactions 
can be efficiently activated, therefore changing the 
average chemical content of the ejecta. With 
very strong mass loss, the HBB nucleosynthesis 
has no time to be completely established. This 
affects 
mainly the oxygen and sodium yields, while it 
leaves almost unaltered the lithium yield. 
In the most massive models a stronger 
mass loss rate (with respect to the standard value
adopted, calibrated on slightly less massive
models to reproduce the lithium-rich stars
luminosity function in the LMC) 
might lead to ejecta which are sodium rich, 
and which show low magnesium isotopic ratios; 
there are two factors behind such findings:

\begin{itemize}
\item{i) Massive AGBs achieve large luminosities
already during the first TPs; if mass loss is
increased during these phases, we have a larger
ejection of material which is still sodium
rich and not extremely $^{24}$Mg depleted.}

\item{ii) As a consequence of the
reduction of the mass of the envelope, the
evolution is halted earlier, before strong sodium and
$^{24}$Mg depletion may take place.}
\end{itemize}
}

\end{enumerate}

For models with masses M$\leq 4.5$M$_{\odot}$ 
the ejecta are sodium rich and with
low magnesium isotopic ratios in any case, with
the only difference that, for larger values
of $\eta_R$, oxygen is scarcely depleted, as
the increase of the temperature at the base of
the external envelope is halted before it may
reach values sufficiently high to activate
efficiently the full CNO cycle.

As a general conclusion, these models show that
the predictive power of AGB models is still
undermined by many uncertainties. The parameters
space, however, has not yet been fully explored
(e.g. the role of extra-mixing at the bottom of
the envelope) and for this reason we should not
discard the hypothesis that massive AGB stars are
responsible for the chemical anomalies observed 
in GCs stars.


\end{document}